\documentclass[journal]{IEEEtran}

\usepackage{algorithm}
\usepackage{algpseudocode}
\usepackage{xurl}
\usepackage{amsmath, balance}
\usepackage{graphicx}
\usepackage{comment}
\usepackage[justification=centering]{caption}
\usepackage{float}
\usepackage{enumitem}

\begin{document}

\title{Metaverse Framework for Wireless Systems Management}

\author{Ilias Chrysovergis, Alexandros-Apostolos A. Boulogeorgos~\IEEEmembership{Senior Member,~IEEE} \\Theodoros A. Tsiftsis,~\IEEEmembership{Senior Member,~IEEE,} and
        Dusit Niyato, \IEEEmembership{Fellow,~IEEE}
\thanks{I. Chrysovergis is with the Department of Informatics \& Telecommunications, University of Thessaly, Lamia 35100, Greece, and also with METATOPIA, Thessaloniki 54624, Greece, E-mail: ichrysovergis@uth.gr}
\thanks{A.-A. A. Boulogeorgos is with the Department of Electrical and Computer Engineering, University of Western Macedonia, ZEP Area, 50100 Kozani, Greece, E-mail: aboulogeorgos@uowm.gr}
\thanks{T. A. Tsiftsis is also with the Department of Informatics \& Telecommunications, University of Thessaly, Lamia 35100, Greece, E-mail: tsiftsis@uth.gr}
\thanks{Dusit Niyato is with the College of Computing \& Data Science (CCDS), Nanyang Technological University, Nanyang Avenue 639798, Singapore, Email:dniyato@ntu.edu.sg}
\thanks{The work of A.-A. A. Boulogeorgos is supported by the project MINOAS. The research project MINOAS is within the H.F.R.I. call ``Basic Research Financing (Horizontal support of all Sciences)" under the National Recovery and Resilience Plan ``Greece 2.0" funded by the European Union - NextGenerationEU (H.F.R.I. Project Number: 15857).}}
   
\maketitle

\begin{abstract}
This article introduces a comprehensive metaverse framework, which is designed for the simulation, emulation, and interaction with wireless systems. 
The proposed framework integrates core metaverse technologies such as extended reality (XR), digital twins (DTs), artificial intelligence (AI), internet of things (IoT), blockchain, and advanced 6G networking solutions to create a dynamic, immersive platform for both system development and management. 
By leveraging XR, users can visualize and engage with complex systems, while DTs enable real-time monitoring and optimization. 
AI generates the three-dimensional (3D) content, enhances decision-making and system performance, whereas IoT devices provide real-time sensor data for boosting the simulation accuracy. 
Additionally, blockchain ensures secure, decentralized interactions, and 5G/6G networks offer the necessary infrastructure for seamless, low-latency communication. 
This framework serves as a robust tool for exploring, developing, and optimizing wireless systems, aiming to provide valuable insights into the future of networked environments.
\end{abstract}


\IEEEpeerreviewmaketitle

\section{Introduction}
\IEEEPARstart{T}{he} rapid advancements in metaverse-related technologies, such as extended reality (XR), digital twins (DT), artificial intelligence (AI), internet of things (IoT), blockchain, and networking, have paved the way for experts across various industries to envision intelligent systems, three-dimensionally-rendered, physics-based, interoperable, secure, and distributed.
These advancements align with the growing complexity of current and next-generation wireless system, which demand more sophisticated tools for visualization, optimization, and management.
The ever-increasing scale and complexity of wireless systems present challenges in terms of resource allocation, performance optimization, and security. 
Traditional network management approaches lack the flexibility and scalability to address the dynamic and multifaceted requirements of these networks. 
As wireless technologies evolve to support AI, ultra-reliable and low-latency communications, immersive communications, integrated sensing and communication, massive communications, and ubiquitous connectivity, the need for innovative solutions becomes critical.

The metaverse offers a transformative approach to these challenges by providing three-dimensional (3D) environments for managing wireless networks. 
By integrating XR, DT, and AI, metaverse-based systems simulate, visualize, and optimize complex network architectures in real-time. 
Operators can predict and mitigate network congestion, simulate failure scenarios, and optimize network coverage in highly dynamic environments.
The upcoming demands of next-generation applications, like autonomous vehicles, smart cities, and industrial automation, necessitate systems that are performant, adaptive, secure, and resilient. 
 Metaverse  bridges the gap between the physical and digital worlds, enabling real-time monitoring and decision-making; thus, transforming next-generation wireless networks from reactive to proactive management paradigms. 

Motivated by the above observations, Lotfi \textit{et al.} \cite{Lotfi2023} addressed the critical issue of incentivizing IoT devices to share sensing data essential for creating accurate DTs within the metaverse. 
Khan \textit{et al.} \cite{Khan2023} explored the potential of metaverse in advancing sixth-generation (6G) wireless systems. 
They presented a metaverse that serves as a collaborative and virtual platform for network designers, developers, and engineers to analyze, optimize, and operate wireless networks. 
Hashash et al.~\cite{hashash2024seven} introduced a novel framework that categorizes metaverse experience into seven distinct worlds: enhanced reality, DTs, mirror worlds, lifelogging, virtual worlds, augmented reality, and mixed reality.
Sehad et al.~\cite{sehad2024generative} presented a framework that explores generative AI (GenAI) to enhance multisensory experiences in next-generation systems. 
This work introduced novel approaches for synthesizing and transmitting sensory data, including haptic, olfactory, and gustatory information, while addressing the unique challenges of bandwidth efficiency and latency requirements.
Khan et al.~\cite{khan2022digital} presented a comprehensive framework that leverages DTs to enable intelligent and self-sustaining 6G systems. 
Their work introduced key design requirements and architectural components for DT implementation in 6G networks. 
The research demonstrated how DTs can create virtual representations of physical 6G systems alongside associated algorithms, enabling advanced simulation capabilities and network optimization. 

Open-radio access network (O-RAN) ALLIANCE's next Generation Research Group (nGRG) has published two significant contributions addressing key aspects of future RAN and their integration with the metaverse ecosystem. 
In~\cite{oran-ngrg-dapps-2024}, the authors explored the integration of distributed apps in RAN environments, establishing critical use cases and technical requirements that enable metaverse applications requiring ultra-low latency and high reliability. 
This work was complemented by their comprehensive study on DT RAN~\cite{oran-ngrg-dtwin-2024}, which investigates the fundamental enablers required to create virtual replicas of RAN infrastructure, a crucial component for building the enterprise metaverse and enabling immersive experiences. 
Polese et al.~\cite{polese2024colosseum} introduced Colosseum, a large-scale wireless network emulator that serves as a DT for O-RAN systems and demonstrated how Colosseum facilitates the testing of complex scenarios involving multiple base stations, user devices, and network slices.

Jiang et al. \cite{jiang2024learnable} articulated an approach using neural representations to create learnable wireless DTs.
Alikhani et al.~\cite{alikhani2024lwm} introduced the large wireless model (LWM), a foundation model designed for wireless channel prediction and characterization.
Yang et al.~\cite{diffusion2024} reported a novel framework that leverages diffusion models to address complex network optimizations in IoT environments. 
Their approach demonstrated that diffusion models can learn high-quality solution distributions, enabling optimal solution generation through repeated sampling.  
Their work presented a novel architecture that learns the underlying physics of wave propagation and can generalize across different wireless environments and frequencies.
Li et al.~\cite{li2024large} proposed a novel framework that leverages LLMs to address multi-objective optimization challenges in UAV-enabled ISAC networks. 
Their work demonstrated how LLMs can effectively balance competing objectives such as communication throughput, sensing accuracy, and energy efficiency while considering various network constraints. 

Zou et al.~\cite{zou2024telecom} introduced TelecomGPT, a framework designed to transform general-purpose LLMs into telecom-specific models. 
Bariah and Debbah \cite{bariah2023ai} investigated the path towards realizing artificial general intelligence (AGI) through AI embodiment, emphasizing the critical role of 6G networks in orchestrating AGI development. 
The authors explored different types of knowledge acquisition and cognition, while highlighting the limitations of auto-regression in large language models. 
Their work demonstrated that sensory grounding through 6G could enhance LLMs' understanding capabilities but alone is insufficient for achieving full human-like comprehension. 
The authors identified that true understanding requires deeper knowledge representation including semantic understanding, causal reasoning, and contextual awareness. 

While the above studies laid the groundwork, they focus on specific facets of the metaverse-wireless nexus. 
A critical gap remains in the literature: the lack of a comprehensive, integrated framework that holistically combines all key enablers, namely XR, DTs, AI, IoT, networking and blockchain, into a unified, multi-layered architecture designed for the end-to-end management of wireless systems. 
Existing approaches either explore these technologies in isolation or propose high-level concepts without detailing a concrete, interoperable structure. 
This fragmentation limits the potential for creating truly cohesive, secure, and intelligent virtual worlds for wireless system management.
Motivated by this, our work introduces a novel framework that not only integrates these disparate technologies but also proposes a detailed blockchain-based architecture to provide a foundational layer of trust, decentralization, and interoperability. 
By spanning from the physical infrastructure to the immersive user interface, our framework aims to provide a complete blueprint for the next generation of network management tools.

The contributions of the paper is summarized as:
\begin{itemize}
    \item We propose a metaverse-based framework that spans from the physical to the application layer, and enables simulation, emulation, and interaction with complex wireless systems.
    \textit{Unlike prior studies that typically focus on these technologies in isolation or present high-level concepts, our work provides a concrete, end-to-end architecture.}
    \item We devise a blockchain architecture to make virtual worlds secure, interoperable, and decentralized.
    \textit{This directly tackles the limitations of centralized platforms by providing a foundational layer of trust and eliminating single points of failure.}
    \item We demonstrate the practical application and effectiveness of the framework through a UAV positioning optimization use case, where a Reinforcement Learning (RL) agent is trained within a DT to maximize network throughput in an urban wireless environment.
    \textit{We move beyond theoretical proposals by providing empirical evidence of the framework's capabilities.}
\end{itemize}

\section{Metaverse Framework}

\begin{figure*}
  \includegraphics[width=\textwidth]{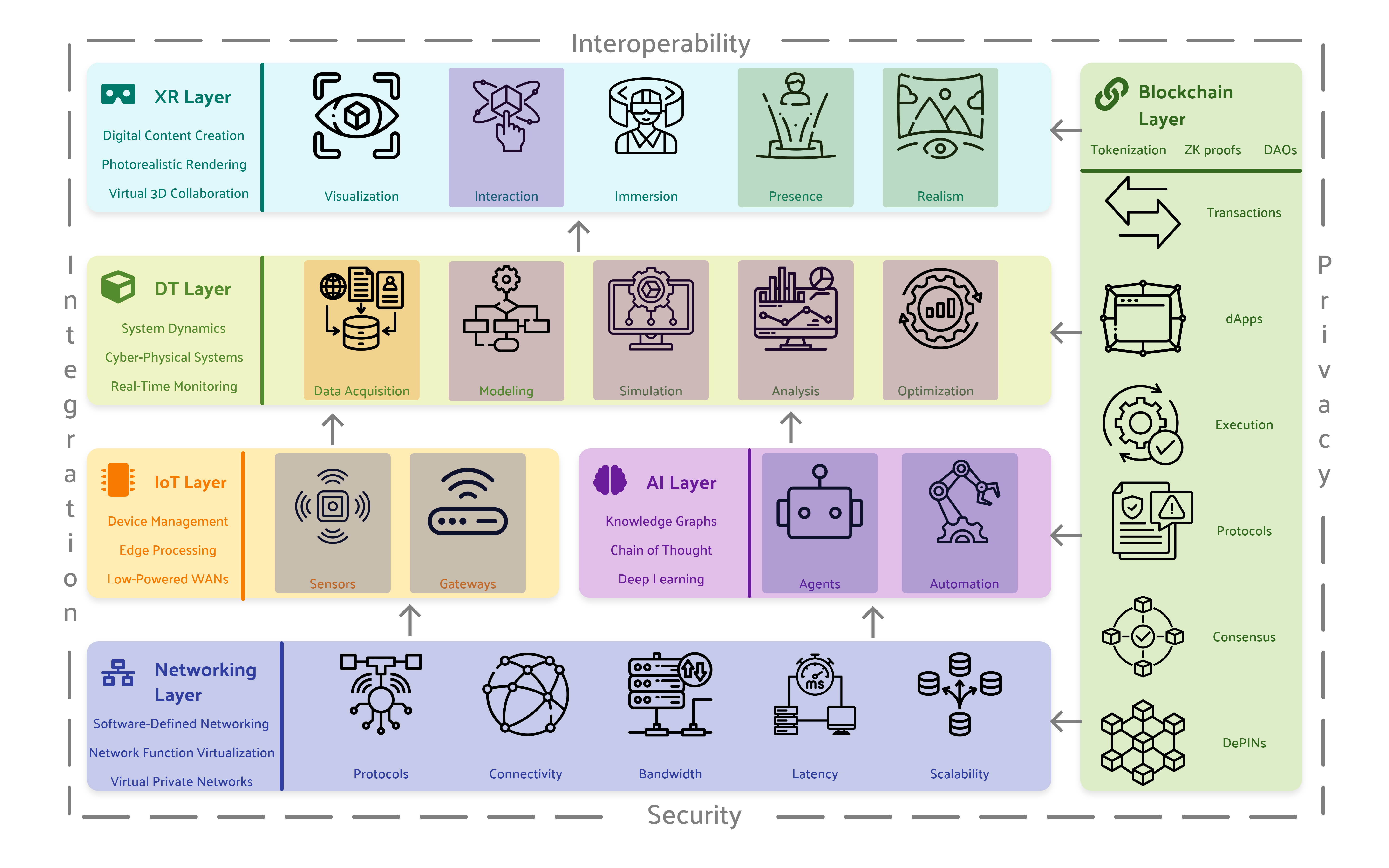}
  \captionsetup{justification=justified, singlelinecheck=false} 
  \caption{
  This multi-layered architecture leverages a foundational Networking Layer, alongside IoT and AI, to feed a central DT layer for advanced simulation and optimization. 
  The results are rendered via an XR layer for immersive interaction, while a vertical Blockchain layer ensures trust, security, and decentralization across the entire system.
  Key principles of security, privacy, and interoperability are enforced across all layers to ensure a robust and integrated system.}
  \label{fig:architecture-components}
\end{figure*}

Figure~\ref{fig:architecture-components} illustrates a multi-layered technology stack that spans from underlying networking infrastructure to an XR interface layer, with blockchain services anchoring trust and privacy. 
The  goal is to showcase how data flows and interoperability are achieved across different system components. 
Each layer has a specific role, outlined as follows:

\paragraph{\textbf{Networking Layer}}
This layer provides the connectivity substrate across the system. 
Key concepts include Software-Defined Networking (SDN), Network Function Virtualization (NFV), and Virtual Private Networks (VPNs), all of which enable flexible and secure management of network resources. 
Additionally, core attributes such as protocols, connectivity, bandwidth, latency, and scalability shape how data moves and how efficiently the system can accommodate growth. 
This layer includes: (i) \textit{high-speed connectivity}, which involves the implementation of 5G networks to support data-intensive applications and real-time interactions, (ii) \textit{network slicing}, which employs techniques to partition the network into virtual slices to ensure quality of service for different applications, and (iii) \textit{edge computing}, which deploys computational resources at the edge of the network to reduce latency and enhance efficiency.
Overall, the networking layer ensures reliable data transport for upper-layer functionalities.

\paragraph{\textbf{IoT Layer}}
The IoT layer comprises physical sensors, gateways, and edge-processing components. 
Device management and wireless access networks are common techniques to handle large-scale deployments and energy constraints. The data collected from environmental sensors, industrial machines, or other real-world entities flows upward for further processing. 
This layer bridges the physical and digital domains.

\paragraph{\textbf{AI Layer}}
It processes incoming data streams, leveraging advanced techniques like knowledge graphs, chain-of-thought reasoning, and deep learning. 
These capabilities facilitate intelligent decision-making and pattern recognition. Modular \emph{agents} autonomously execute tasks, while \emph{automation} extends these agents’ capabilities into real-world or virtual control. 
The AI layer adds cognitive and analytical depth, transforming raw sensor signals into actionable insights leading to optimal solutions.

\paragraph{\textbf{DT Layer}}
DTs create virtual replicas of physical systems, enabling real-time monitoring, diagnostics, and augmentation. 
The DT layer consists of the following components: 
(i) \textit{Data acquisition}, which involves integration with IoT devices to gather real-time data from the physical system, (ii) \textit{modeling and simulation}, which includes algorithms for creating and updating digital replicas based on real-time data, (iii) \textit{analysis and optimization}, which provides tools for analyzing system performance and augmenting operations using the DT, and (iv) \textit{visualization}, which utilizes platforms for rendering 3D scenes and user interfaces of the digital system.

\paragraph{\textbf{XR Layer}}
It acts as the human-facing interface that visualizes and interacts with DTs and AI-driven algorithms. 
Tools include digital content creation, photorealistic rendering, and collaborative virtual environments. 
Core attributes—visualization, interaction, immersion, presence, and realism—help deliver seamless user experiences, enabling stakeholders to explore data-rich scenarios, train in simulated worlds, or collaborate on designs in real time.
The layer is responsible for rendering 3D models of components, and providing an interactive user interface. 
It supports the following functionalities: (i) \textit{3D visualization}, which is used for real-time rendering of the system components and the components spatial relations, (ii) \textit{interactive controls} that are composed by the user interfaces for manipulating and configuration components within the virtual environment, and (iii) \textit{simulation scenarios} composed by predefined and customizable scenarios for training and testing AI models.

\paragraph{\textbf{Blockchain Layer}}
It ensures decentralized trust and secure transactions across this complex ecosystem. 
Methods like tokenization, zero-knowledge proofs, and decentralized autonomous organizations (DAOs) belong to this layer. 
It provides mechanisms for immutable record-keeping, consensus, and privacy, which are  critical when data and digital assets are verified or exchanged among multiple, potentially untrusted entities.
This layer ensures secure, decentralized, and interoperable interactions within the metaverse ecosystem. 
Key components include: (i) \textit{Security}, which leverages cryptography to ensure all operations are recorded immutably on the distributed ledger, (ii) \textit{interoperability} that uses smart contracts to facilitate seamless data sharing and collaboration across various networks, and (iii) \textit{decentralization}, that ensures that the blockchain's distributed nature prevents any single entity from controlling the system; thus, promoting fairness and transparency. 

\paragraph{\textbf{Cross-Cutting Aspects}}
The framework emphasizes seamless integration and interoperability among all layers. 
This is achieved through standardized interfaces, data models, and communication protocols. 
Throughout the stack, \emph{integration} underscores how each layer builds upon data and services from the one below. 
The integration layer ensures that data flows smoothly between XR, DTs, AI, IoT, blockchain, and networking components, providing a unified platform for simulating, emulating, managing, and interacting with wireless systems. 
\emph{Interoperability} highlights the importance of seamless communication across disparate systems. 
Examples include an XR front-end that pulls DT data protected by blockchain, or AI models that might depend on real-time sensor data processed at the edge. 
 \emph{Security and privacy} are enforced at every layer to safeguard data authenticity, system resilience, and user trust.
Security and privacy are critical considerations in this framework. 
The security layer implements robust mechanisms to protect data integrity, confidentiality, and availability. 
This includes: (i) \textit{Authentication and authorization}, which employs JSON web tokens for secure and stateless user authentication, (ii) \textit{data encryption}, which ensures that data is protected in transit and at rest to guard against unauthorized access, (iii) \textit{access control}, which implements fine-grained policies to restrict data and resource access based on user roles and permissions, and (iv) \textit{password-less systems}, which utilize modern authentication methods such as biometrics, email or SMS-based one-time codes, or hardware security keys (e.g., FIDO2) to replace traditional password-based authentication, enhancing both security and user convenience.

\section{Blockchain-based Decentralized Architecture for Simulating Wireless Systems}

Blockchain technology plays a pivotal role in enhancing the metaverse capabilities, particularly by enabling a secure, interoperable, and decentralized ecosystem. 
One of the key integrations of blockchain is the establishment of a token-based economy, where users can interact with virtual assets, services, and resources. 
This economy built on blockchain, ensures that all transactions are tamper-proof. 
Users are able to purchase tokens that allow them to access computing resources, storage, and data processing capabilities. 
These tokens can be used to obtain virtual equipment, data sets, or custom-built simulations. 
Blockchain’s tokenization capabilities provide a framework for managing ownership and access to these digital assets, allowing users to buy, sell, or lease them securely within the marketplace. 
For example, a company could tokenize a custom-built DT of a real communication system and offer it to other companies for testing in exchange for tokens. 

Blockchain enables the automation of resource management through the use of smart contracts. 
These self-executing contracts ensure that once a user pays for a service or resource using tokens, the required computing power, storage, or simulation capabilities are automatically allocated.
Smart contracts enforce the terms of the transaction, such as usage limits or access periods, without needing manual intervention. 
This automation improves efficiency, allowing users to focus on system optimization and experimentation while the blockchain handles resource distribution. 
Another use of blockchain is to secure data flows between IoT devices and the virtual environment.
Blockchain’s decentralized structure ensures that no single entity controls the entire system, reducing vulnerabilities and increasing trust.
Moreover, blockchain ensures that sensitive data related to industrial processes, network configurations, or real-time monitoring is protected.
In the marketplace, blockchain technology enables peer-to-peer exchange of digital assets. 
Users can directly trade virtual equipment, AI models, or simulation tools using tokens. 
Blockchain ensures that all ownership transfers are immutable, fostering a trustless environment where users can collaborate and innovate freely.

\subsection{Architecture}

\begin{figure}[h]
    \centering
    \includegraphics[width=\columnwidth]{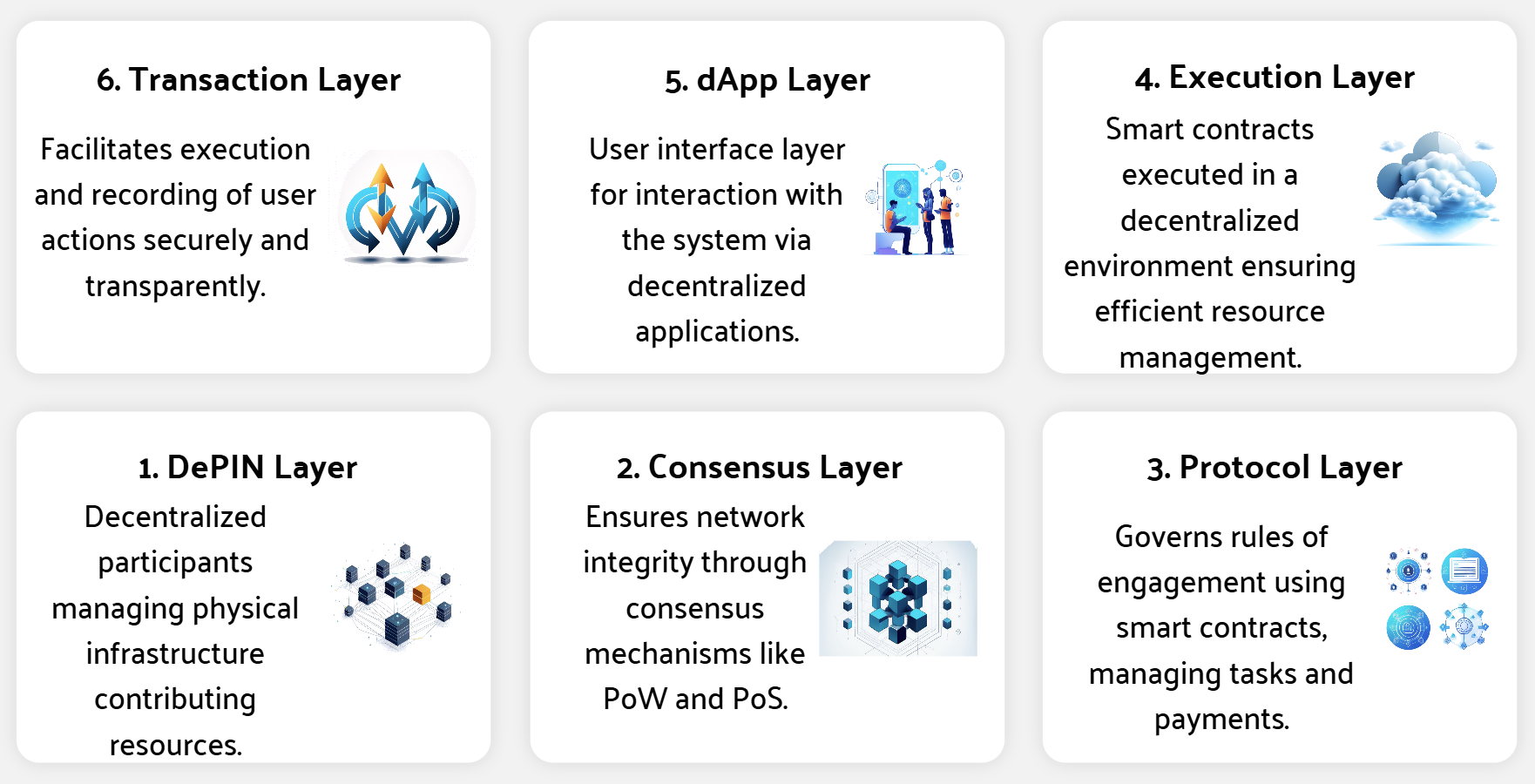}
    \caption{Blockchain-based decentralized framework for simulating DTs.}
    \label{fig:blockchain_framwork}
\end{figure}

The architecture (Fig.\ref{fig:blockchain_framwork}) leverages decentralization at every step, from the physical infrastructure level to the final user interaction, ensuring trustless coordination and transparency across all layers of the system. 
The sublayers of the architecture are the following:

\subsubsection{\textbf{DePIN Sublayer}}
The blockchain's architecture begins with the DePIN Sublayer, where decentralized participants manage the physical infrastructure that contributes data, bandwidth, or services. 
These participants operate nodes in the network, providing the necessary computational or storage resources that power decentralized applications (dApps) or network services. 
The infrastructure's contribution is logged and validated via a blockchain mechanism, ensuring decentralization, transparency, and rewards for node operators.

\subsubsection{\textbf{Consensus Sublayer}}
Moving up to the Consensus Sublayer, the architecture employs a decentralized ledger that uses consensus mechanisms such as Proof of Work (PoW) or Proof of Stake (PoS) to validate transactions and ensure network integrity. 
Each transaction, including those requesting computation from a DePIN node, is validated through this consensus mechanism. 
This layer ensures that all nodes in the network agree on the state of the system, ensuring the integrity and security of each computation or service request.

\subsubsection{\textbf{Protocol Sublayer}}
At this sublayer, smart contracts govern the rules of engagement, managing how tasks are assigned to DePINs, handling payment via escrow, and ensuring that the nodes comply with agreed-upon service terms. 
These smart contracts automate processes such as computation task assignment, payment distribution, and validation of results.
Additionally, in cases where a DAO is in place, decisions about the system’s governance can be managed collectively by token holders through this layer.

\subsubsection{\textbf{Execution Sublayer}}
In this sublayer, the smart contracts are executed. 
The actual computation takes place. 
The selected DePIN performs the computation task requested by the user, and the results are submitted back to the blockchain. 
During execution, gas costs manage resource consumption, ensuring the network's efficiency and preventing abuse.

\subsubsection{\textbf{dApp Sublayer}}
At this sublayer, users interact with the system through a front-end interface, connecting them to the blockchain through wallets and smart contracts. 
The dApp provides a user-friendly environment where the computation request is initiated, the result is displayed, and transactions are handled transparently. 
This layer is critical for user engagement and the overall usability of the decentralized system.

\subsubsection{\textbf{Transaction Sublayer}}
It facilitates the execution and recording of user actions, such as submitting tokens or provisioning services.
When a user requests a computation via the dApp, the transaction is broadcasted to the blockchain, validated, executed by a DePIN node, and finalized. 
The transaction moves through all relevant layers, including validation and result submission, and concludes when the result and payment are confirmed on the blockchain. 
The user receives the final result after transaction finality, ensuring a secure, transparent, and decentralized system.

\subsection{Decentralized computation flow}

In the decentralized computation flow, the process begins at the transaction layer, where a user requests a computation via a decentralized application (dApp). 
This request is signed and broadcasted to the blockchain as a transaction. 
At the consensus layer, the transaction is validated by the network using a consensus mechanism, ensuring that it meets all requirements. 
Following validation, a suitable DePIN node is automatically selected to carry-out the computation. 
The task is then managed by the protocol layer, where smart contracts handle the assignment, escrow the payment for the service, and ensure that the DePIN adheres to the task requirements.

Once the DePIN node has completed the computation, it submits the results to the blockchain via the execution layer. 
The protocol layer verifies the accuracy of the computation and, if all conditions are met, releases the payment to the DePIN node. 
The transaction and computation are then finalized in the consensus layer, ensuring that both the result and payment are permanently recorded on the blockchain. 
Finally, at the transaction layer, the user is notified of the result through the dApp, receiving the outcome of the computation after the transaction has reached finality on the blockchain. This structured process ensures a decentralized, secure, and transparent approach to computation tasks.

\section{UAV Positioning Optimization for Throughput Maximization}

\begin{figure*}
    \centering    \includegraphics[width=\textwidth]{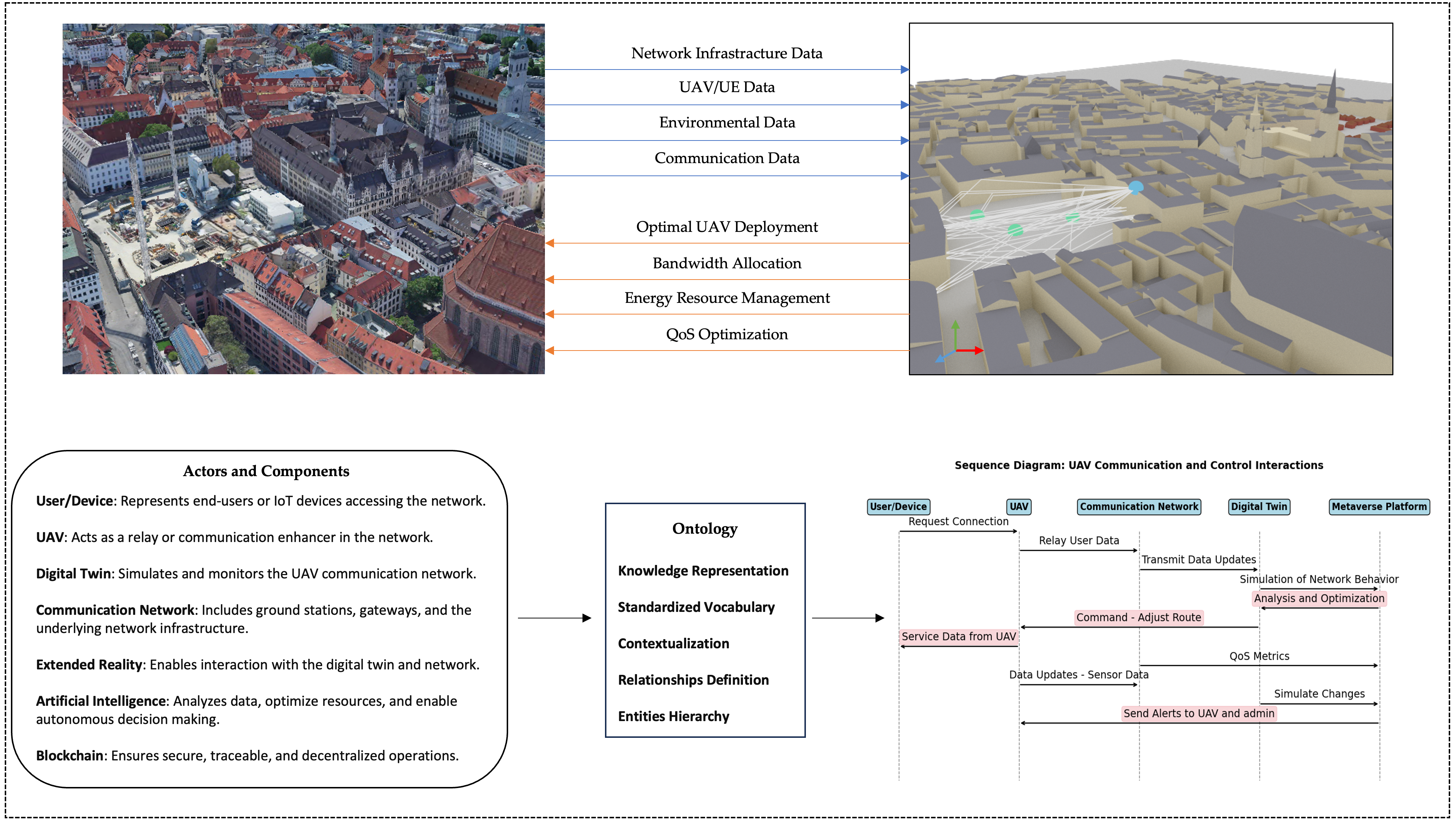}
    \caption{UAV positioning optimization for throughput maximization.}
    \label{fig:uav_rl_simulation}
\end{figure*}

To demonstrate the key capabilities of the proposed metaverse framework, this section demonstrates a use case focused on the AI-driven optimization of a UAV's position. 
While the complete architecture envisions the integration of all layers, including XR and blockchain, this study provides a proof-of-concept implementation of the core intelligence. 
It leverages a high-fidelity DT to train an RL agent to enhance network throughput in a dynamic urban environment.

\subsection{Envisioned System Architecture and Interaction Flow}

Figure~\ref{fig:uav_rl_simulation} presents the conceptual architecture and operational flow of the UAV positioning use case. 
The framework establishes a seamless link between a physical urban environment and its corresponding DT. 
This digital replica is central to the envisioned system, serving as the environment where the UAV's operations can be modeled, analyzed, and optimized before being deployed in the real world.

In this architecture, data from the real world—including network infrastructure status, UAV and user equipment telemetry, environmental conditions, and communication metrics—are continuously feeding the DT. 
This influx of real-time information would ensure that the virtual representation remains synchronized with its physical counterpart, enabling accurate simulations. 
The AI layer processes the data to generate actionable intelligence, leading to optimized outputs such as ideal UAV deployment strategies, dynamic bandwidth allocation, efficient energy resource management, and enhanced Quality of Service (QoS).

The system is composed of several key actors and components:
(a) \textbf{User/Device:} The end-point in the network, representing ground-based users or IoT devices that require reliable connectivity.
(b) \textbf{UAV:} An airborne communication relay, dynamically positioned to enhance signal strength and network throughput for ground users.
(c) \textbf{DT:} A virtual, real-time model of the UAV, the network, and the surrounding environment. 
It is the primary tool for simulation and monitoring.
(d) \textbf{Network:} The terrestrial infrastructure, including base stations and core network components, that connects the UAV and users to the wider network.
(e) \textbf{XR:} The interface that allows human operators to visualize the network state and interact with the DT in an immersive 3D space.
(f) \textbf{AI:} The core intelligence of the system. 
Its algorithms analyze data from the DT to drive the decision-making process for optimization.
(g) \textbf{Blockchain:} Underpins the system by ensuring that all operations, data exchanges, and transactions are secure, traceable, and decentralized.

The ontology is the conceptual backbone that enables the coherent operation of the entire framework. 
It provides a formal, structured knowledge representation by establishing a standardized vocabulary and defining the entities, their properties, and the permissible relationships between them.
The primary function of the ontology is to ensure semantic interoperability. 
It creates a common language that allows heterogeneous components, such as the physical UAV hardware, the software-based AI models, and the distributed blockchain ledger, to unambiguously understand each other's state, data, and commands. 
For example, the ontology formally defines what a `UAV` is, what attributes it possesses (e.g., `position`, `batteryLevel`), and what actions it can perform (e.g., `relayData`, `updatePosition`).

By providing the logical blueprint, the ontology models the system's complex interactions. 
The sequence diagram shown in Fig.~\ref{fig:uav_rl_simulation} is a direct visual representation of a valid operational flow, derived from the rules and relationships defined within the ontology. 
In essence, the ontology transforms a collection of independent components into a cohesive, intelligent system whose behavior can be precisely simulated and executed.

The sequence diagram illustrates the dynamic data-driven workflow for optimizing the UAV's position.
The process begins when a device requests a connection, which the UAV relays to the ground network.
The network transmits all relevant operational data, such as the connection request and current signal metrics, to the DT, ensuring the virtual model remains synchronized with the physical world. 
The DT forwards this real-time data to the central metaverse platform. 
AI algorithms analyze the current network state, user locations, and other environmental factors to compute an optimal adjustment for the UAV, like a new flight trajectory or hovering position, to maximize throughput.
The platform sends the optimized command back to the DT. 
The twin then translates this virtual command into an actionable instruction for the physical UAV, which adjusts its position accordingly.

Throughout this process, a continuous feedback loop is maintained. 
The UAV and network send back QoS metrics (e.g. throughput, BER, delay, SINR) and sensor data, while external systems provide supplementary information like weather updates. 
The platform uses this data to refine its simulations, visualize the network state in real-time, and send alerts if necessary. 
This interactive and data-driven workflow allows the system to autonomously optimize the UAV's position to maximize communication throughput for users on the ground.

\subsection{AI-Driven Optimization and Implementation}
The core of this use case is the implementation of an RL agent tasked with identifying an optimal positioning policy for the UAV. 
The problem is framed as an RL task, where an intelligent agent learns to make sequential decisions in a simulated environment to maximize a cumulative reward.
The agent's goal is to identify the best possible location for the UAV that maximizes the sum of the signal-to-interference-plus-noise ratio (SINR) for all ground receivers, which  correlates with the overall network throughput.

The implementation leverages the proximal policy optimization (PPO) algorithm, a state-of-the-art RL technique, to train the agent. 
The learning environment is a custom DT built using the Sionna \cite{hoydis2022sionna} ray-tracing (RT) engine, which provides a realistic simulation of radio environment in an urban landscape.
The starting positions of the UAV and the receivers are chosen to represent a realistic urban scenario where the receivers are spread out on the ground (at a height of 1.5 meters, typical for user devices), and the UAV starts in a position that requires optimization to improve signal quality.
The state observed by the agent at each step is the UAV's current 3D coordinates. 
The action space is discrete, allowing the agent to move the UAV incrementally along the x, y, or z axes. 
Following each action, the environment calculates the resulting SINR for each receiver, and the sum of these values is returned to the agent as a reward signal, guiding its learning process toward optimal positioning strategies.
During the training phase, the PPO agent interacts with the DT over multiple episodes, iteratively refining its policy through trial-and-error to achieve convergence on high-reward states that correspond to superior UAV positions.
The entire training workflow is detailed in Algorithm~\ref{alg:uav_optimization}.

The time complexity of the algorithm is dominated by its nested loops, running for $E$ episodes each with $T$ steps, resulting in a base complexity of $O(E\,T)$. 
In each step, key operations include state observation and action selection via a MLP policy at $O(H \, W^2)$, where $H$ and $W$ are the number of hidden layers of the MLP and their width, respectively, position updates at $O(1)$, channel impulse response computation at $O(R \, P \, L)$ with $R$ being the receivers, $P$ the paths, and $L$ the scene complexity, SINR calculation at $O(R \, M)$, where $M$ represents the number of multipath components, reward computation at $O(R)$, transition storage at $O(1)$, and policy updates at $O(H \, W^2)$. $L$ quantifies the scene complexity, typically representing the number of geometric elements or objects (e.g., triangles, polygons) within the simulated 3D environment that the RT process accounts for.
The total time complexity is $O(E \, T \, (R \, P \, L + H \, W^2))$, with RT being the primary bottleneck in urban scenarios.

The space complexity is  driven by the storage requirements of the simulation environment, including the urban scene's 3D mesh data and RT buffers at $O(L + R)$, the PPO agent's MLP weights at $O(H \cdot W^2)$, and episode transition buffers at $O(T \, D)$ where $D$ represents the dimensionality of a single transition tuple. 
The total space complexity is $O(L + H \, W^2 + E \, T \, D)$, though peak usage per episode is $O(L + H \, W^2 + T \, D)$ since transitions are not persisted across episodes, with the scene complexity $L$ typically dominating in memory-intensive urban simulations.

\begin{algorithm}
\caption{RL-based UAV Positioning for Throughput Maximization}
\label{alg:uav_optimization}
\begin{algorithmic}[1]
\State \textbf{Initializion Phase:} Create the simulation environment using Sionna, loading the urban scene (`munich`).
Define one UAV `Transmitter` and three `Receiver` devices with initial positions.
Instantiate the PPO agent with a multi-layer perceptron (MLP) policy.
\For{each training episode}
    \State Reset the environment, placing the UAV and receivers at their starting locations.
    \For{each step $t$ in the episode}
        \State Observe the current state $s_t$ (UAV's 3D position).
        \State The PPO agent selects an action $a_t$ (a discrete movement vector) based on its current policy $\pi(a_t|s_t)$.
        \State Apply the action to the UAV, updating its position in the simulation.
        \State Use the Sionna ray-tracer to compute the channel impulse responses between the UAV and all receivers.
        \State Calculate the SINR for each receiver based on the channel conditions.
        \State Compute the reward $r_t$ as the sum of all receiver SINRs.
        \State The PPO agent stores the transition $(s_t, a_t, r_t, s_{t+1})$.
        \State Update the agent's policy using the collected transition data to maximize the expected cumulative reward.
    \EndFor
\EndFor
\State \textbf{Output:} A trained PPO agent with an optimized policy for UAV positioning.
\end{algorithmic}
\end{algorithm}

\subsection{Analysis of Training Results}
The efficacy of the RL agent in optimizing the UAV's position is demonstrated by the training results, which are presented as plots of the average SINR and the resulting channel capacity over 300 training episodes. 
These metrics illustrate the agent's learning progress and the direct impact of its policy on network performance.

\begin{figure}
    \centering
    \includegraphics[width=\columnwidth]{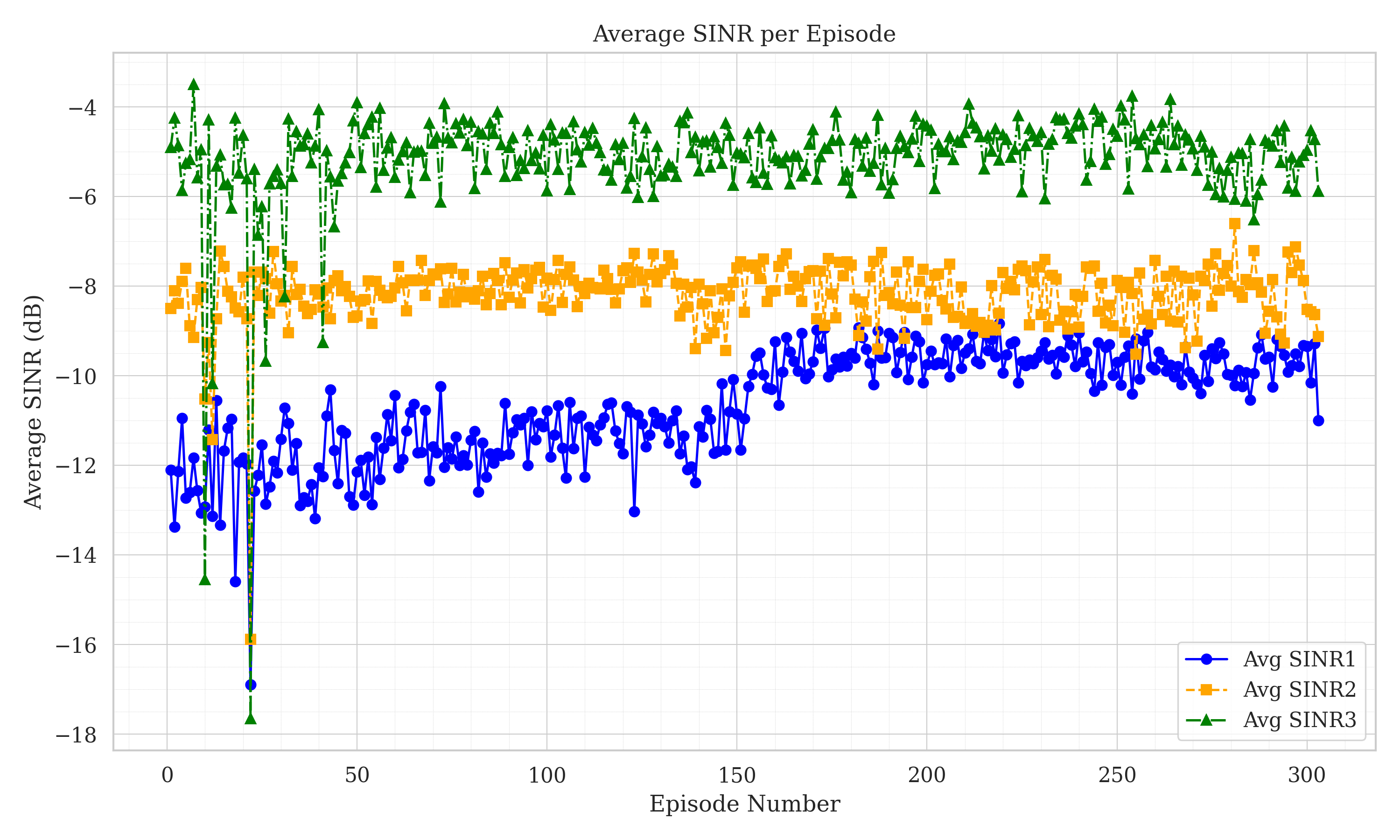}
    \caption{Average SINR per Episode during RL Training.}
    \label{fig:sinr_results}
\end{figure}

\begin{figure}
    \centering
    \includegraphics[width=\columnwidth]{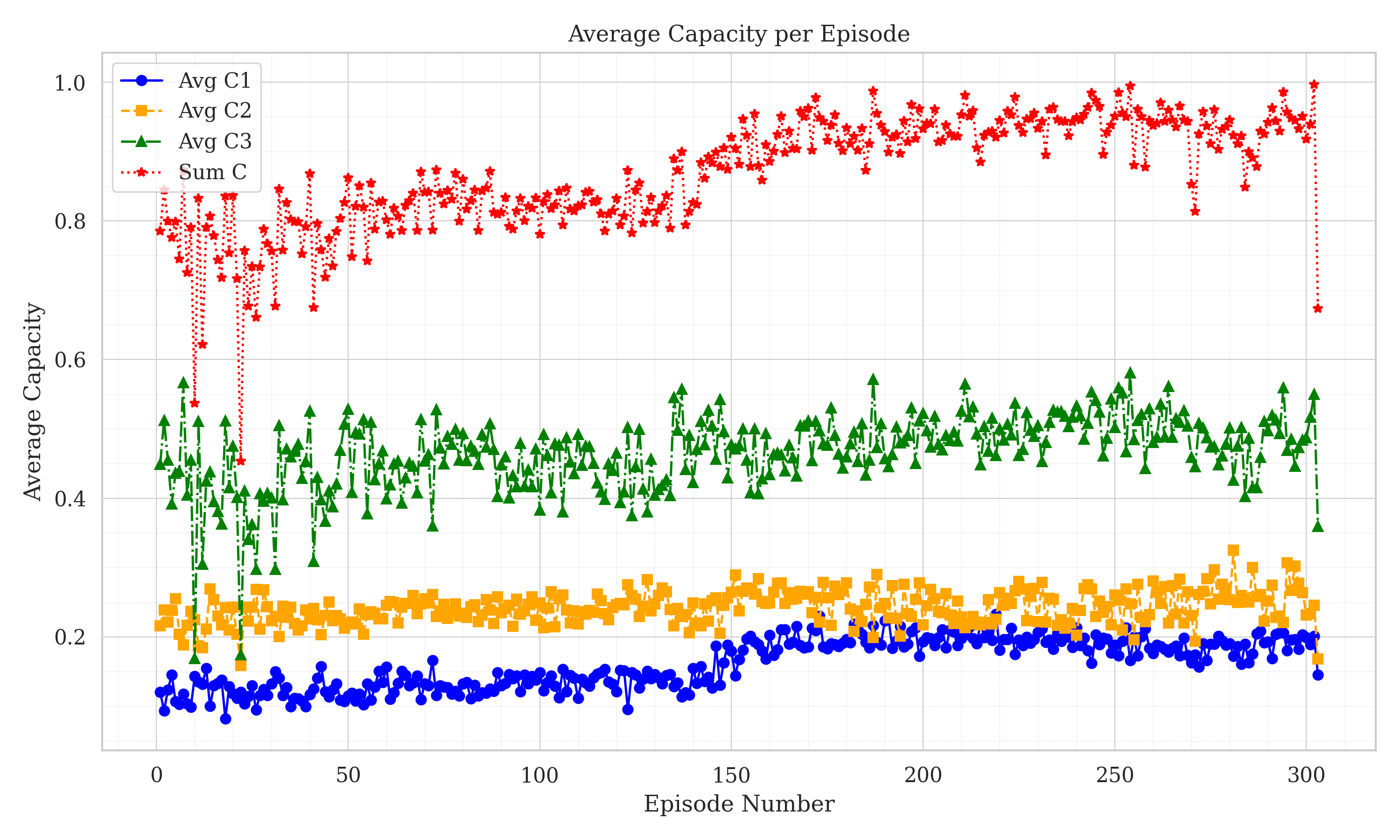}
    \caption{Average Capacity per Episode during RL Training.}
    \label{fig:capacity_results}
\end{figure}

Figure~\ref{fig:sinr_results} shows the average SINR for each of the three ground receivers. 
The figure clearly indicates a successful learning trend. 
Initially, the SINR values are low and/or volatile, particularly for receiver 1 (blue line), which starts at approximately -12 dB.
As the training progresses, the agent discovers more effective UAV positions. 
A significant improvement is visible around episode 150, after which the SINR for receiver 1 stabilize at a higher level around -10 dB.
This demonstrates that the agent has successfully learned a policy that improves the link quality for the less advantageous user, without reducing the QoS of the other users.
The direct consequence of this SINR optimization is shown in Fig.~\ref{fig:capacity_results}, which plots the average channel capacity for each receiver and their sum. 
The total network capacity (red line) exhibits a distinct learning curve that mirrors the SINR improvements, rising significantly and then stabilizing at a much higher level after episode 150. 
This confirms that the agent's learned policy not only achieves its abstract goal of maximizing SINR but also translates this into a tangible and significant increase in the overall data throughput of the wireless network.

\section{Conclusion \& Future Directions}

The proposed metaverse framework suggests a significant advancement in how we simulate, emulate, and interact with wireless systems.
By integrating XR, DTs, AI, IoT, blockchain and advanced networking technologies, immersive virtual worlds, can be created. 
The modular and scalable design of the framework, provides a versatile tool for addressing the complexities of modern systems.
The use of DTs allows for real-time monitoring and optimization of network performance, while AI-driven algorithms provide solutions for dynamic network management.
The integration of IoT devices ensures that the system can adapt to real-world conditions, enhancing the accuracy and reliability of simulations.
Moreover, the XR layer offers an interactive interface for users, making complex network data more accessible and easier to manage.
By leveraging these technologies, the proposed architecture offers a novel tool for network operators, researchers, and developers to explore new technologies, optimize existing systems, and enhance their understanding of networks.
In summary, the metaverse framework represents a significant step forward in the analysis and synthesis of wireless systems.

There are several areas for further research. 
\begin{itemize}[leftmargin=*, label=\textbullet, itemsep=5pt]
    \item \textbf{GenAI for Virtual World Creation:} A significant area for future work is leveraging GenAI for content creation. 
    By using generative models to produce 3D assets, textures, and interactive elements, researchers and developers can rapidly prototype and iterate on dynamic, evolving virtual worlds.
    \item \textbf{Deepening O-RAN Integration and Advancement:} Future research on Open RAN integration within the framework will focus on several critical areas:
    \begin{itemize}[leftmargin=*, label=\textendash, itemsep=3pt]
        \item \textbf{High-Fidelity DTs:} Developing specialized and accurate DTs for O-RAN components, including O-RU, O-DU, and O-CU elements, to enable real-time simulation of complex network behaviors.
        \item \textbf{Specialized AI/ML for RICs:} Creating enhanced AI/ML capabilities tailored for the near-real-time and non-real-time RAN Intelligent Controllers (RICs) to facilitate more sophisticated network optimization and automation.
        \item \textbf{Support for Advanced O-RAN Features:} Evolving the architecture to support functionalities like network slicing through dedicated dApp implementations and smart contracts within the blockchain layer.
        \item \textbf{Real-Time Synchronization:} Improving synchronization mechanisms between the physical O-RAN infrastructure and its DT to support mission-critical, ultra-low-latency applications.
        \item \textbf{Semantic Communication:} Integrating semantic communication principles to enhance the efficiency of data exchange between physical and virtual network elements while adhering to O-RAN standards.
    \end{itemize}
    \item \textbf{Comprehensive Parametric Analysis:} Building upon the proof-of-concept of this paper, a crucial next step is to conduct a comprehensive parametric analysis of the UAV positioning use case. 
    This involves systematically varying key system parameters to fully characterize the framework's performance, scalability, and robustness under a wide range of diverse operational conditions.
\end{itemize}

\balance

\end{document}